\documentclass[11pt]{article}
\usepackage{moriond,epsfig}

\def\Journal#1#2#3#4{{#1} {\bf #2}, #3 (#4)}

\def\NPB{{\em Nucl. Phys.} B}

\def\PRL{\em Phys. Rev. Lett.}

\def\EPJ{{\em Eur. Phys.} J}

\def\be{\begin{equation}}
\def\ee{\end{equation}}
\def\bea{\begin{eqnarray}}
\def\eea{\end{eqnarray}}

\begin{document}
\vspace*{4cm}
\title{W and Z (Plus Jets) Production and Asymmetries at 1.96~TeV}

\author{ Gavin Hesketh, on behalf of the CDF and D\O\ Collaborations }

\address{Northeastern University, Boston, MA, 02115, USA}

\maketitle\abstracts{
This paper reviews the latest results from the Tevatron on W and Z physics, and their use as probes of QCD.
}

\section{W and Z Physics at the Tevatron}\label{subsec:intro}

Many of the physics aims at the Tevatron and LHC, such as top physics, the W mass measurement and Higgs searches rely heavily upon the understanding of underlying QCD processes.
The W and Z bosons are well understood, and can be used as ``standard candles'' to study QCD and jet production with high precision.

Leptonic W and Z decay modes provide clean signals with low backgrounds, and high statistics samples are available in Run II.
Production rates and properties of the W and Z are precisely predicted within the Standard Model (SM), and this is one of the few areas in which NNLO theory can be tested to high precision. 
The W charge asymmetry and Z rapidity are sensitive to PDFs, and can be used to constrain PDF uncertainties.
Studies of the production of associated jets with the W and Z allow a test of perturbative QCD, and are important for tuning simulation at the Tevatron and LHC. 
W/Z plus jets are also the main background to top physics, Higgs searches and many searches for new phenomena.
The W and Z signals are also used to improve the understanding of detector performance, and form the basis of any physics analysis using high energy leptons.

\section{Inclusive W and Z Production Cross Sections}
The benchmark of understanding W and Z production at a hadron collider is a precise measurement of the inclusive cross section multiplied by leptonic branching fraction ($\sigma\times$BR).
This is a test of the NNLO theory prediction \cite{wzcross} and of the understanding of detector performance.
The approach is simple. 
Event selection is based upon triggering on and reconstructing the high energy leptons, with neutrinos being identified by the presence of high missing transverse energy (MET).
Detector efficiencies are measured in data, and the acceptance is calculated with Pythia Monte Carlo\cite{Pythia}.
Backgrounds are dominated by QCD processes: semi-leptonic quark decays; and hadronic jets misidentified as electrons.
These are reduced by requiring the leptons to be separated from any hadronic activity in the event. 
Other main backgrounds come from Z$\rightarrow\tau\tau$ and W$\rightarrow\tau\nu$\ in which the taus decay to electrons or muons; and in the case of W$\rightarrow\mu\nu$, Z$\rightarrow\mu\mu$\ where one muon is not reconstructed. 
These contributions are estimated using Monte Carlo.

Measurements of these cross sections have been presented in previous years\cite{cross_sec}, and two updated results from the CDF collaboration are presented here.
First, a measurement of Z$\rightarrow\mu\mu$, using 337~pb$^{-1}$~of data. 
9620 Z candidates are selected by requiring each muon to have a p$_{\mathrm{T}} > 20$~GeV, and the invariant mass of the pair to lie between 66 and 116~GeV.
The result obtained is $\sigma\times\mathrm{BR} = 261.2 \pm 2.7 (stat) ^{+5.8}_{-6.1} (syst) \pm 15.1 (lum)$~pb, compared to the theoretical prediction of $251.3 \pm 5.0$~pb\cite{cdf_wz}.
Second, a measurement of W$\rightarrow \mathrm{e}\nu$ using 223~pb$^{-1}$\ of data.
Event selection is based upon electrons in the range $1.2<|\eta|<2.8$, requiring the electron E$_{\mathrm{T}} > 20$~GeV and MET~$> 25$~GeV.
The result obtained is $\sigma\times\mathrm{BR} = 2815 \pm 13 (stat) ^{+94}_{-89} (syst) \pm 169(lum)$~pb, which is in good agreement with the result obtained with central electrons ($|\eta| < 1$), proving the ability of CDF to reconstruct electrons in forward rapidity regions.
This can be compared to the theoretical prediction of $2687 \pm 54$~pb\cite{cdf_wz}.

\begin{figure}[h]
\begin{center}
\psfig{figure=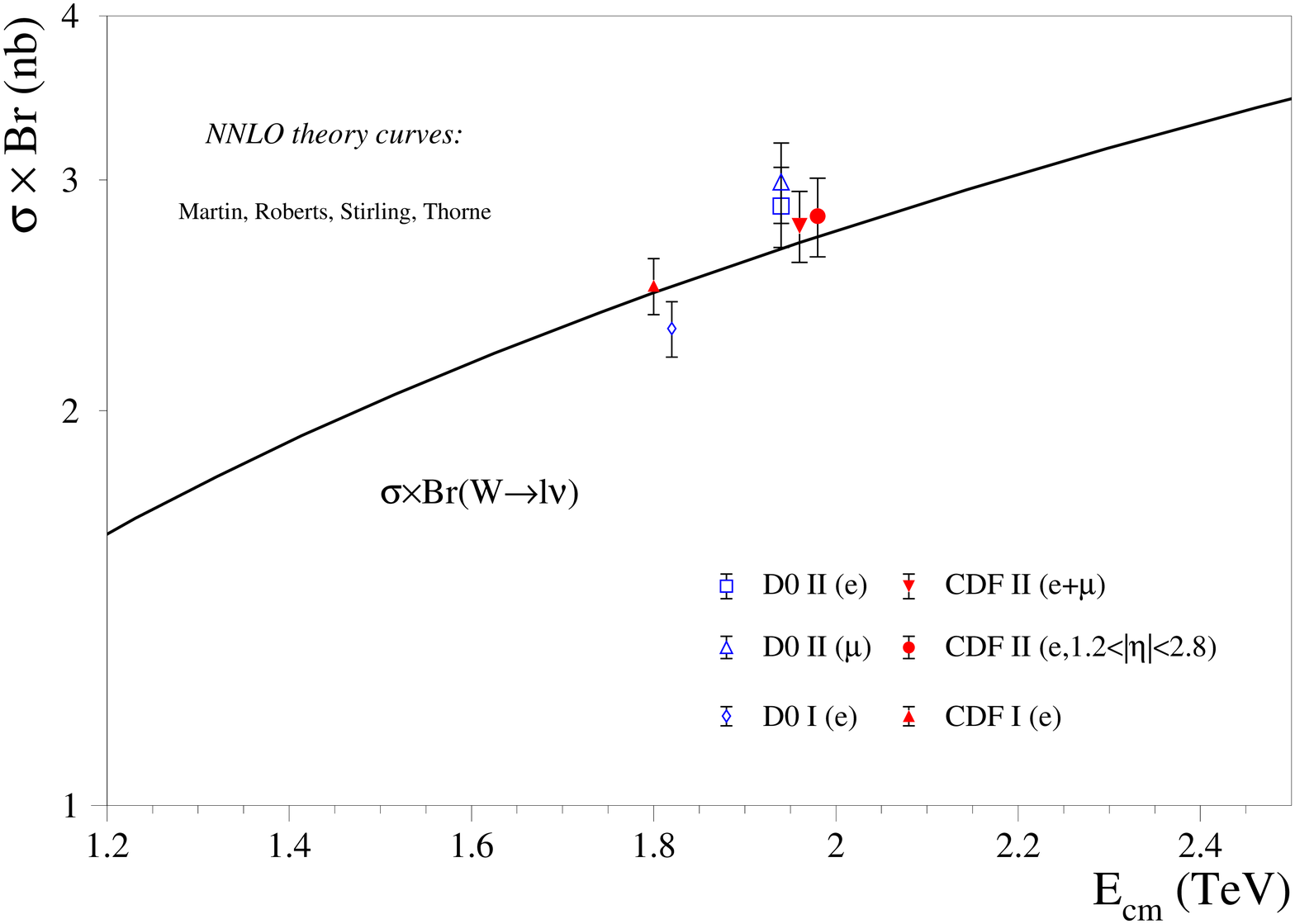,height=1.7in}
\psfig{figure=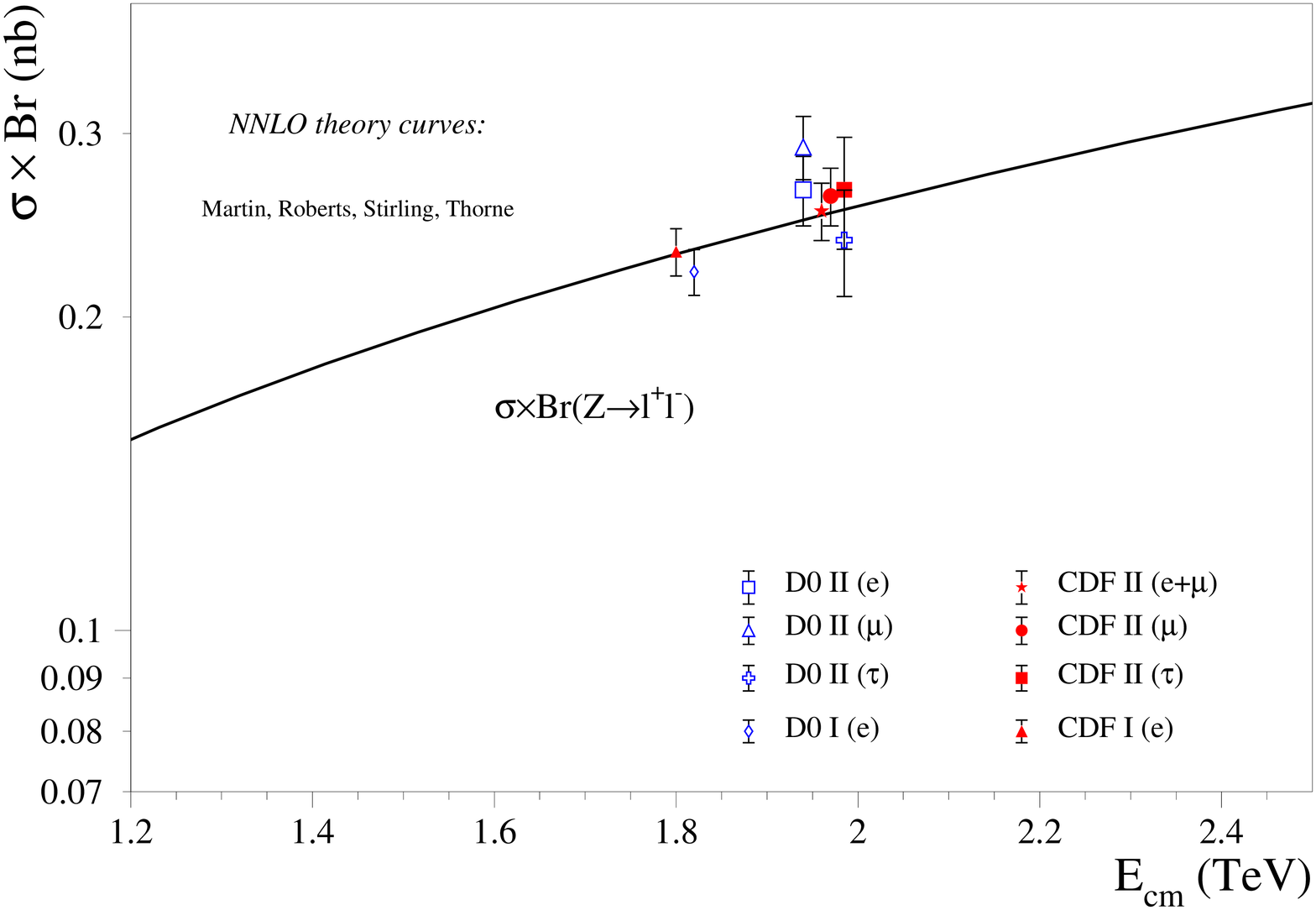,height=1.7in}
\end{center}
\caption{Summary plots for inclusive W (left) and Z (right) production cross sections times leptonic branching fractions, compared to the NNLO theory curves.} 
\label{fig:cross_sec}
\end{figure}

All Tevatron Run~II inclusive cross section measurements are in agreement with the NNLO SM predictions, and are summarised in figure \ref{fig:cross_sec}, including previously published results from CDF\cite{cdf_wz}.
Limiting systematics include the uncertainties on lepton identification (1-2\%), variations in acceptance due to PDF uncertainties (1-2 \%), and the uncertainty on the total luminosity (6\% for CDF, 6.5\% for D\O).

\section{W Boson Charge Asymmetry}
As we have seen, W and Z production, as well as many other measurements, are affected by the understanding of PDFs.
Measurement of the W$^+$\ to W$^-$\ production asymmetry in proton anti-proton collisions is directly sensitive to differences in the u and d quark PDFs, and a measurement of this asymmetry can be used to constrain these PDFs.
The presence of a neutrino in leptonic W decays complicates W reconstruction, so instead the charged lepton asymmetry as a function of rapidity, $A(\eta_l)$, is measured.
This is a convolution of the W asymmetry and the V-A decay, and factors approximately into the $u$\ and $d$\ PDFs:
\begin{equation}
A(\eta_l) = \frac{d\sigma(l^+)/d\eta - d\sigma(l^-)/d\eta}{d\sigma(l^+)/d\eta + d\sigma(l^-)/d\eta} \simeq \frac{d(x)}{u(x)}
\end{equation}
D\O\ presents a new result in the muon channel, based on 230~pb$^{-1}$. 
The asymmetry is most sensitive to PDFs at high $|\eta|$, so this measurement benefits from the ability of D\O\ to trigger on and reconstruct muons out to $|\eta| < 2$.
Event selection is based on one muon with p$_{\mathrm{T}} > 25$~GeV and MET $>$\ 25~GeV. 
Cuts are placed on the muon track quality to control the charge misidentification rate, which is below 0.01 \%.
The obtained asymmetry is show in figure \ref{fig:asym}, and compared to the predictions from the MRST\cite{MRST} and CTEQ\cite{CTEQ} PDF sets, including the spread due to the CTEQ PDF error sets.
The asymmetry is limited by statistics in the high $\eta$\ region, so with the addition of more data it will be possible to constrain these PDF error sets.

\section{Z Boson Rapidity}
\begin{figure}[h]
\begin{center}
\psfig{figure=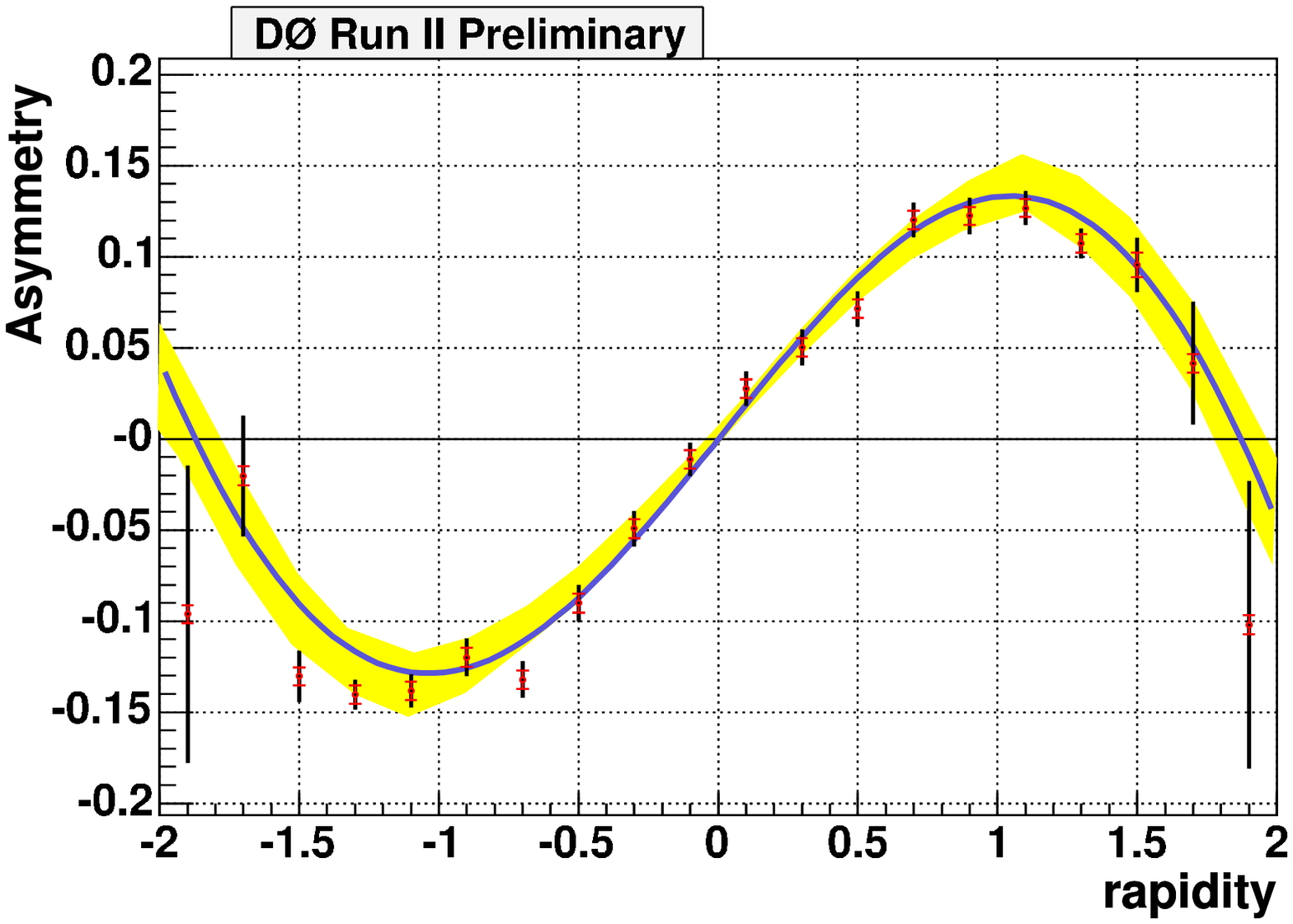,height=1.7in}
\psfig{figure=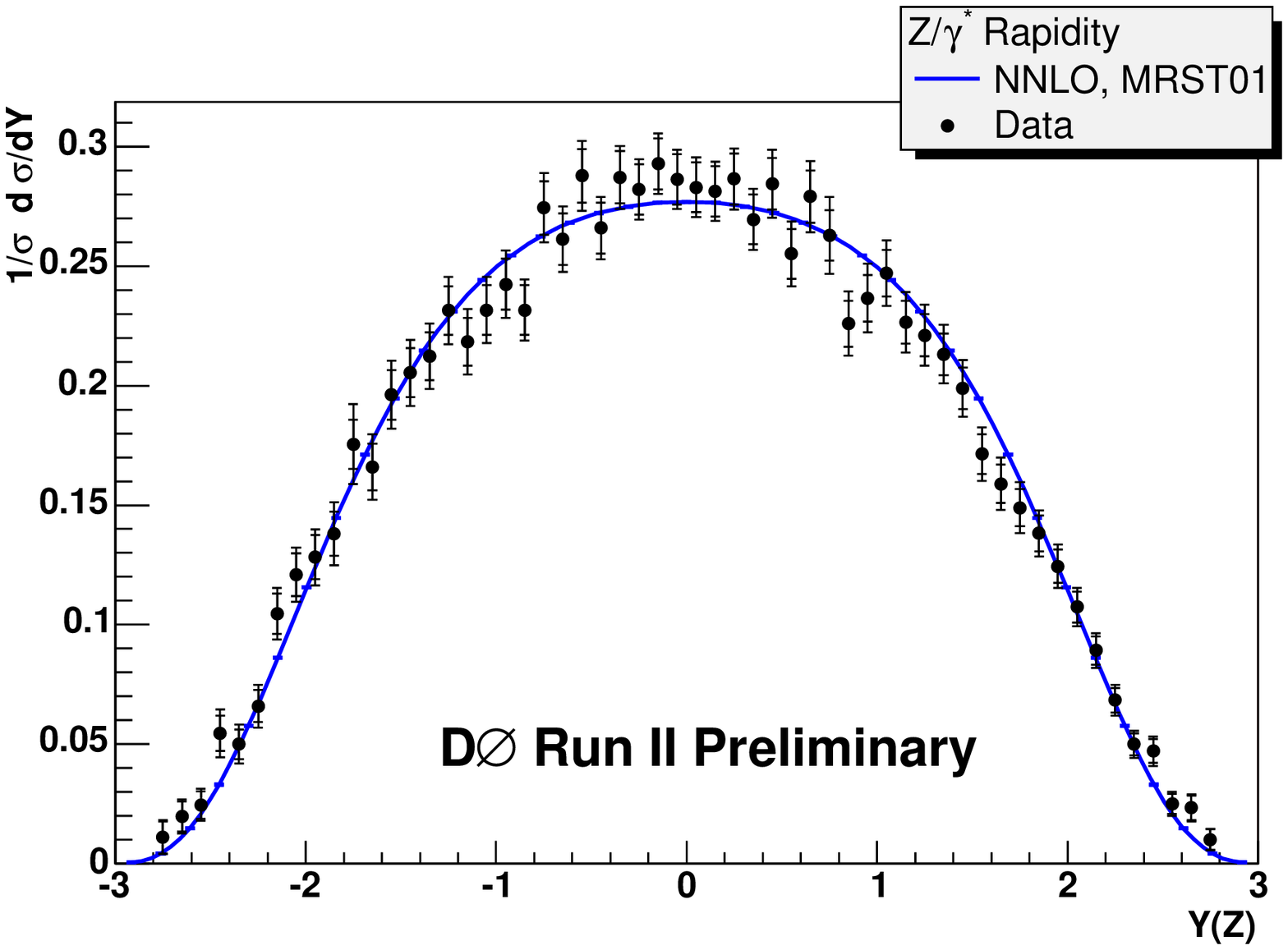,height=1.7in}
\end{center}
\caption{The left plot shows the W lepton charge asymmetry as a function of rapidity, with data points shown with statistical (black) and systematic (red) uncertainties, and compared to the prediction from the MRST PDF set (blue line) and the CTEQ PDF set and error sets (yellow band).
The right plot shows the Z rapidity distribution, showing the data points compared to NNLO theory prediction using the MRST PDF set.
\label{fig:asym}}
\end{figure}

The Z rapidity distribution is also sensitive to PDFs. 
Compared to the W charge asymmetry, there is the benefit that the Z can be fully reconstructed, but it is not possible to separate the u and d PDFs.
D\O\ presents a new result on the Z rapidity in the electron channel, based on 337~pb$^{-1}$. 
Figure \ref{fig:asym} shows the differential cross section as a function of Z rapidity, comparing data to the NNLO prediction using the MRST PDF sets.
It can be seen that there is good agreement between the theory and data.

\section{W and Z Plus Jets Production}
The production of W and Z in association with jets is a crucial test of perturbative QCD. 
The signals of these events are also very similar to top quark production and production of the Standard Model Higgs boson in association with a W or Z (the main search channels at the Tevatron).
A good understanding of W and Z plus jets therefore affects many of the Tevatron physics aims. 
In the LHC era, understanding of higher jet multiplicities will be even more important.

Several Monte Carlos of boson plus jet production exist.
Most use a combination of perturbative QCD matrix element calculation and parton showering to produce the jet spectra seen in data. 
Tevatron results are vital to the tuning of these simulations.

In order to allow unbiased comparisons to theory, the experimental aim is to perform a model-independent analysis. 
Leptons and jets are limited to measured and well understood regions with cuts on $\eta$\ and p$_{\mathrm{T}}$, and there is no extrapolation outside these regions. 
The reconstructed jets are then unsmeared to the particle level. 

\begin{figure}[h]
\begin{center}
\psfig{figure=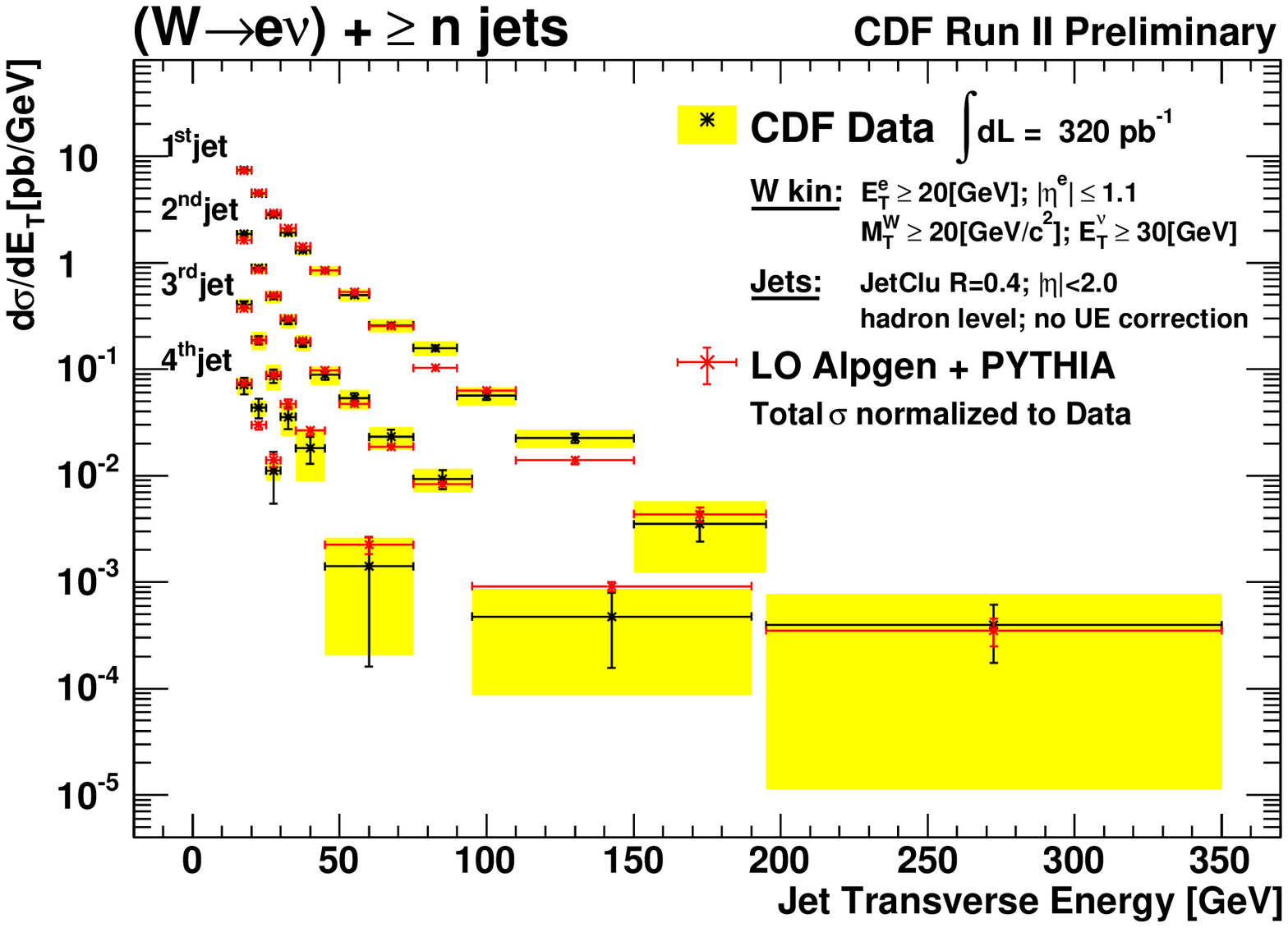,height=2in}
\psfig{figure=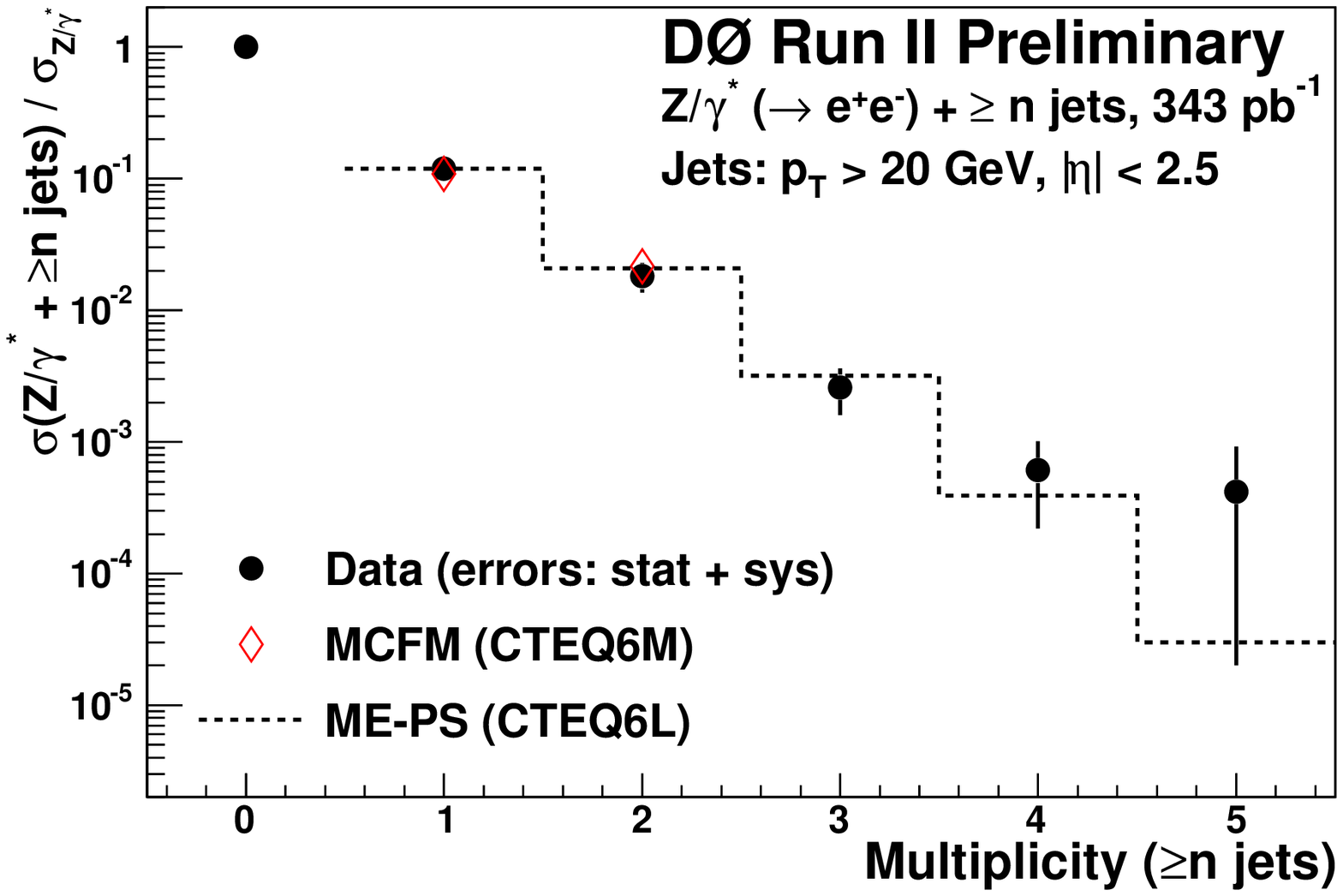,height=2in}
\end{center}
\caption{Comparison of jet E$_\mathrm{T}$\ in W plus jet events between CDF data and ALPGEN prediction (left). Comparison of jet multiplicities in Z plus jet events between D\O\ data and MCFM and MADGRAPH + Pythia.
\label{fig:VBjets}}
\end{figure}

CDF presents an updated analysis of W plus jets in the electron mode, based on 320 pb$^{-1}$. 
The W selection is based upon one electron with $|\eta| < 1.1$\ and p$_{\mathrm{T}} > 20$~GeV, MET$ > 30$~GeV and a reconstructed transverse mass$ > 20$~GeV. 
Jet finding is done with the JETCLU algorithm with R=0.4, in the range $|\eta| < 2$. 
The data are compared to ALPGEN v2\cite{ALPGEN}, a leading matrix element Monte Carlo, using Pythia for parton showering.
As ALPGEN is leading order only, it does not correctly predict the inclusive cross section, so the Monte Carlo and data are normalized to the same number of events in each inclusive jet multiplicity bin.
Figure \ref{fig:VBjets} shows the jet transverse energy distributions for each jet, showing good data - Monte Carlo agreement up to 350~GeV.

D\O\ presents an analysis of Z$\rightarrow$ee plus jets based on 343 pb$^{-1}$. 
The event selection is based on two electrons with $|\eta| < 1.1$\ and E$_\mathrm{T} > 25$~GeV.
Jets are identified using a cone algorithm with R=0.5, requiring E$_\mathrm{T} > 20$~GeV and $|\eta| < 2.5$.
Figure \ref{fig:VBjets} shows a comparison between the unsmeared data and ALPGEN, with Pythia used for parton showering. 
Also shown are the inclusive jet multiplicities in data compared to two predictions from Monte Carlo, using MCFM\cite{MCFM}, a NLO Monte Carlo, and MADGRAPH\cite{MADGRAPH} tree level with Pythia for parton showering.
Good agreement between data and Monte Carlo is seen. 

Finally, D\O\ presents a detector-level comparison of Z$\rightarrow$ee plus jets to Sherpa Monte Carlo\cite{SHERPA}, based on 0.95 fb$^{-1}$. 
Sherpa does both matrix element and parton showering, using the CKKW\cite{CKKW} matching scheme. 
Sherpa successfully describes jet multiplicities seen in data, as well as all jet properties, such as kinematic and angular distributions. 
Sherpa therefore proves to be a very useful tool for future analyses.


\section*{References}
\begin{thebibliography}{99}
\bibitem{wzcross}C.R. Hamberg, W.L. van Neerven and W.B. Kilgore, \Journal{\NPB} {359}{343}{1991}.
\bibitem{Pythia} T. Sjostrand, \Journal{\em JHEP}{05}{026}{2006} .
\bibitem{cdf_wz}D. Acosta \it{et. al.}, (CDF Collaboration), \Journal{\PRL}{94}{091803}{2005}.
\bibitem{cross_sec}A. Bellavance, hep-ex/0506025.
\bibitem{MRST}A. Martin, R. Roberts, W. Sterling and R. Thorne, \Journal{\EPJ}{4}{463}{1998}.
\bibitem{CTEQ}J. Pumplin \it{et.al}., \Journal{\em JHEP}{0207}{012}{2002}.
\bibitem{ALPGEN} M.L. Mangano \it{et. al.}, \Journal{\em JHEP}{0307}{001}{2003}.
\bibitem{MCFM}See the website http://mcfm.fnal.gov/
\bibitem{MADGRAPH} F. Maltoni and T. Stelzer, \Journal{\em JHEP}{0302}{027}{2003}.
\bibitem{SHERPA}Gleisberg \it{et. al.}, \Journal{\em JHEP}{0402}{056}{2004}.
\bibitem{CKKW}F. Krauss, \Journal{\em JHEP}{0208}{015}{2002}.

\end{thebibliography}
\end{document}